\documentclass{llncs}
\usepackage{booktabs}
\usepackage[utf8]{inputenc}
\usepackage{amssymb}
\usepackage[table,xcdraw]{xcolor}
\usepackage{graphicx}
\usepackage{xspace}
\usepackage{amsmath}
\usepackage{url}      

\begin{document}
\title{A matter of words: \\NLP for quality evaluation of Wikipedia medical articles}
\subtitle{\footnotesize{An extended abstract of this work will appear as a short paper in proceedings of ICWE2016, published by LNCS Springer.}}
\titlerunning{A matter of words}  
%
\author{Vittoria Cozza\inst{1} \and Marinella Petrocchi\inst{1}
\and Angelo Spognardi\inst{2}}
\authorrunning{Cozza et al.} 
%
\tocauthor{Vittoria Cozza, Marinella Petrocchi, Angelo Spognardi}
\institute{IIT CNR, Pisa, Italy
\email{\{v.cozza, m.petrocchi\}@iit.cnr.it}
\and
DTU Lingby, Denmark
\email{angsp@dtu.dk}
}

\maketitle              

\begin{abstract}
  Automatic quality evaluation of Web information is a task with many fields of
  applications and of great relevance, especially in critical domains   like the medical one. We move from the intuition that 
  the quality of content of medical Web documents is affected by features related with the specific domain.  First, the usage of a specific vocabulary
  (Domain Informativeness); then, the adoption of specific codes (like those
  used in the infoboxes of Wikipedia articles) and the type of document
  (e.g., historical  and technical ones). In this paper, we propose to
  leverage specific domain features to improve the results of the
  evaluation of Wikipedia medical articles. In particular, we evaluate the articles 
   adopting an ``actionable" model, whose
  features are related to the content of the articles, so that the model can
  also directly suggest strategies for improving a given article quality.  We
  rely on Natural Language Processing (NLP) and dictionaries-based techniques in order to extract the bio-medical
  concepts in a text. We prove the effectiveness of our approach by classifying the medical articles of the Wikipedia Medicine
  Portal, which have been previously manually labeled by the Wiki Project team. The results of
  our experiments confirm that,  by considering domain-oriented features, it
  is possible to obtain sensible improvements with respect to existing
  solutions, mainly for those articles that other approaches have less correctly classified. 
  Other than being interesting by their own, the results call for further research in the area of domain
  specific features suitable for Web data quality assessment.
\end{abstract}

%
%
%


\section{Introduction}
As observed by a recent article of Nature News~\cite{Nature2015}, ``Wikipedia is among the most frequently visited websites in the world and one of the most popular places to tap into the world's scientific and medical information". Despite the huge amount of consultations, open issues still  threaten a fully confident fruition of the popular online open encyclopedia. 

A first issue relates to the reliability of the information available: 
since Wikipedia can be edited by anyone, regardless of their level of expertise, this tends to erode the average reputation of the sources, and, consequently, the trustworthiness of the contents posted by those sources. In an attempt to fix this shortcoming,   Wikipedia has recently enlisted the help of scientists to actively support the editing on Wikipedia~\cite{Nature2015}. Furthermore, lack of control may lead to the publication of fake Wikipedia pages, which distort the information by inserting, e.g., promotional articles and promotional external links. Fighting vandalism is one of the main goals of the Wikimedia Foundation, the nonprofit organization that supports Wikipedia: machine learning techniques have been considered to offer a service to ``judge whether an edit was made in good faith or not"~\cite{technologyreview}. 
Nonetheless, in the past recent time, malicious organisations have acted disruptively with purposes of extortion - see, e.g., the recent news on the uncovering of a blackmail network of accounts, which threatened celebrities with the menace of inserting offending information on their Wikipedia pages\footnote{\url{https://en.wikipedia.org/wiki/Wikipedia:Long-term_abuse/Orangemoody}}.


Secondly, articles may suffer from readability issues: achieving a syntactical accuracy that helps the reader with a fluid reading experience is ---quite obviously--- a property which articles should fulfill. 
Traditionally, the literature has widely adopted well known criteria, as the ``Flesch-Kincaid" measure"~\cite{Kincaid1975}, to automatically assess readability in textual documents. More recently, new techniques have been proposed too, for assessing the readability of natural languages (see, e.g., \cite{orletta2014} for the Italian use case, \cite{Sjoholm2012} for the Swedish one, \cite{VajjalaMeurers2014} for English). 

In this paper, we face the quest for quality assessment  of a Wikipedia article, in an automatic way that comprehends not only readability and reliability criteria,  but also additional parameters testifying completeness of information and coherence with the content one expects from an article dealing with specific topics, plus sufficient insights for the reader to elaborate further on some argument. The notion of data quality we deal with in the paper is coherent with the one suggested by recent contributions (see, e.g.,~\cite{Pasi13}), which points out like the quality of Web information is strictly connected to the scope for which one needs such  information.


Our intuition is that groups of articles related to a specific topic and falling within specific scopes are intrinsically different from other groups on different topics within different scopes. We approach the article evaluation through machine learning techniques. Such techniques are not new to be employed for automatic evaluation of articles quality. As an example, the work in~\cite{Wang2013} exploits classification techniques based on structural and linguistic features of an article.  Here, we enrich that model with novel features that are domain-specific. As a running scenario, we focus on the Wikipedia medical portal. Indeed, facing the problems of information quality and ensuring high and correct  levels of informativeness  is even more demanding when health aspects are involved. Recent statistics report that Internet users are increasingly searching the Web for health information, by consulting search engines, social networks, and specialised health portals, like that of Wikipedia. As pointed out by the 2014 Eurobarometer survey on European citizens' digital health literacy\footnote{\url{http://ec.europa.eu/public_opinion/flash/fl_404_sum_en.pdf}}, around six out of ten respondents have used the Internet to search for health-related information. This means that, although the trend in digital health literacy is growing, there is also a demand for a qualified source where people can ask and find medical information which, to an extent, can provide the same level of familiarity and guarantees as those given by a doctor or a health professional. 

We anticipate here that leveraging new domain-specific features is in line with this
demand of articles quality. Moreover, as the outcomes of our
experiments show, they effectively improve the
classification results in the hard task of multi-class assessment,
especially for those classes that  other automatic approaches worst classify.  Remarkably,  our proposal is general enough to be
easily extended to other domains, in addition to the medical one.

Section~\ref{sec:dataset} first describes the structure of the articles present in the medical portal. Then, it gives  details on the real data used in the experiments, which are indeed articles extracted from the medical portal and labeled according to the manual assessment by  the Wikimedia project.
Section~\ref{sec:baseline} briefly presents the actionable model in~\cite{Wang2013}: we adopt it as the baseline for our analysis. In Section~\ref{sec:medical}, we present the domain-specific, medical model we newly adopt in this paper as an extension of the baseline. The extended model includes features specifically extracted from the medical domain. One novel feature is based on the article textual content. Section~\ref{sec:nlp} presents the process which its extraction relies on, with a non trivial analysis of natural language and domain knowledge. 
Section~\ref{sec:exp} presents experiments and results, with a comparison of the baseline model with the new one. In Section~\ref{sec:RW}, we survey related work in the area and in Section~\ref{sec:concl} we conclude the paper.

\section{Dataset}\label{sec:dataset}

\begin{figure}[ht]
\centering
\includegraphics[width=\textwidth]{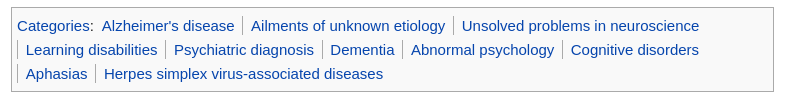}
\caption{Example of Wikipedia Medicine Portal article categories\label{fig:cat}}
\end{figure}

We consider the dataset consisting of the
entire collection of articles of the Wikipedia Medicine Portal,
updated at the end of 2014. 
Wikipedia articles are written according to the Media Wiki markup language, a HTML-like language.
Among the structural elements of one page, which differs from standard HTML pages, there are \textit{i)} the internal links, i.e.,  links to other Wikipedia pages, different from links to external resources); \textit{ii)}  categories, which represent the Media Wiki categories a page belongs to: they are encoded in the part of text within the Media Wiki ``categories" tag in the page source, 
and \textit{iii)} informative boxes, so called ``infoboxes", which summarize in a structured manner some peculiar pieces of information related the topic of the article.
%
%
%
The category values for the articles in the medical portal  span over the ones listed at \url{https://en.wikipedia.org/wiki/Portal:Medicine}. Examples of categories, which appear at the bottom of each Wikipedia page, are in Fig.~\ref{fig:cat}. 

\begin{figure}
\centering
\includegraphics[scale=0.2]{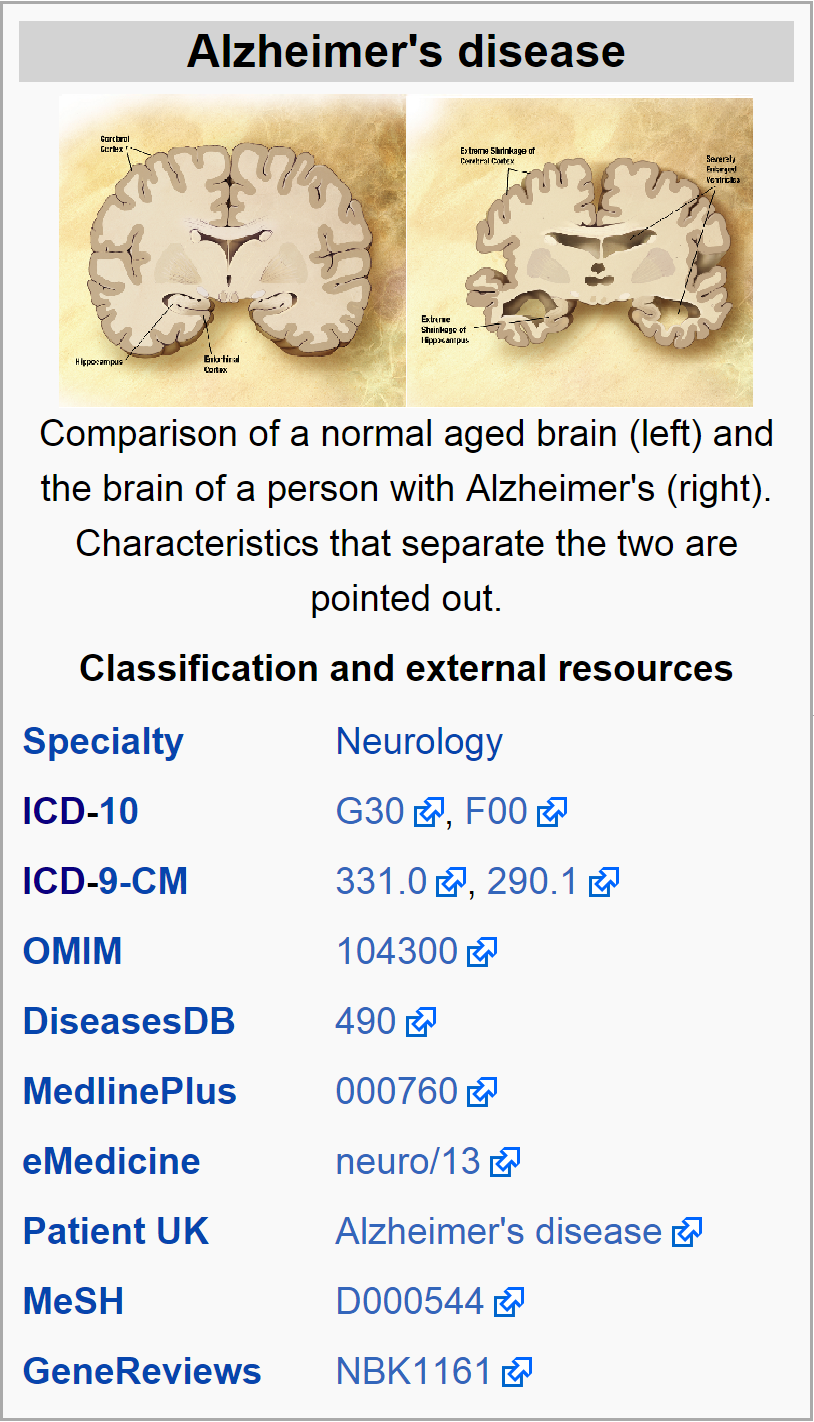}
\caption{The infobox on Alzheimer's disease}
\label{fig:infobox}
\end{figure}

Infoboxes of the medical portal feature medical content and standard coding. As an example, Fig.~\ref{fig:infobox} shows the infobox in the Alzheimer's disease page of the portal. The infobox contains explanatory figures and text denoting peculiar characteristics of the disease and the  value for the standard code of such disease (ICD9, as for the international classification of the disease\footnote{\url{http://www.who.int/classifications/icd/en/}}). 

Thanks to WikiProject
Medicine\footnote{\url{https://en.wikipedia.org/wiki/Wikipedia:WikiProject_Medicine/Assessment}}, 
the dataset of articles we collected from the Wikipedia Medicine Portal has been manually labeled into seven quality classes. 
They
are ordered as  \textit{Stub, Start,
  C, B, A, Good Article (GA), Featured Article (FA)}. The Featured and
Good article classes are the highest ones:  to have those labels, an article  requires a
community consensus and an official review by selected editors, while the other labels can
be achieved with reviews from a larger, even controlled, set of editors. 
Actually, none of the articles in the dataset is labeled as \textit{A}, thus, in the following, we do not consider that class, restricting the investigation to six classes. 

At the date of our study, we were able to gather 24,362 rated
documents. Remarkably, only a small percentage  of them (1\%) is labeled as \textit{GA} and \textit{FA}. Indeed, 
 the distribution of the
articles among the classes is highly skewed. There are very few (201) articles
for the highest quality classes (FA and GA), while the vast majority
(19,108) belongs to the lowest quality ones (Stub and Start). This holds not only for the medical portal. Indeed, it 
is common in all Wikipedia, where, on average, only one article in
every thousand is a Featured one.  

In Section~\ref{sec:exp}, we will adopt a set of machine-learning classifiers to automatically label the articles into the quality classes. 
Dealing with imbalanced classes is a common situation in many real
applications of classification learning: healthy patients over the
population, fraudulent actions over daily genuine transactions, and so
on.  Without any countermeasure, common classifiers tend to correctly
identify only articles belonging to the majority classes, clearly
leading to severe mis-classification of the minority classes, since
typical learning algorithms strive to maximize the overall prediction
accuracy.
%
To reduce the disequilibrium among the size of the classes, we have first randomly sampled the articles belonging to the most populated classes.
Then, we have performed some further elaboration, as shown in the following.

Many studies have been conducted to improve learning algorithms
accuracy in presence of imbalanced data~\cite{imbalancedData}.
For the current work, we have considered one of the most popular approaches, namely
the Synthetic Sampling with Data Generation, detailed
in~\cite{chawla}. It consists in generating synthetic instances from
the minority classes, to balance the overall dataset. The approach has been broadly
applied to problems relying on NLP
features, see, e.g.,~\cite{Chawla:2004:ESI:1007730.1007733}.  In our case, we
resampled the input data set by applying the Synthetic Minority
Oversampling TEchnique (SMOTE\footnote{Implemented and available in
  the Weka framework}), with percentage 40\% for GA and 180\%, for FA. 
In particular, the
steps to oversample are the following:
\begin{itemize}
\item New instances are generated using as seed real examples from the
  minority class;
\item For each real example, its $k$ ($k=5$) nearest neighbours
  examples are identified;
\item Synthetic instances are generated to be at a random point
  between the seed and the neighbours.
\end{itemize}
%
%

 Table~\ref{table:t1} shows the number of articles in the dataset, divided per class, as well as the random samples we have considered for our study. The experiments presented in Section~\ref{sec:exp} are based on the articles of the right-hand column in the table. 
\begin{table}[htb]
\centering
\setlength{\tabcolsep}{12pt}
\begin{tabular}{lccc}
\toprule
~    &	~	& \textbf{with majority} &	 \textbf{with minority} \\
\textbf{class}&\textbf{original dataset} & \textbf{classes sampling}& \textbf{classes oversampling}\\
\midrule
Stub &9,267&1,015&1,015\\
Start & 9,841&1,015&1,015\\
C &3,149&1,015&1,015\\
B &1,894&1,015&1,015\\
GA &153&153&214\\
FA &58&58&162\\
\midrule
total&24,362&4,271&4,436\\
\bottomrule
\end{tabular}
\caption{Dataset{\label{table:t1}}}
\end{table}

\section{Baseline: the actionable model}\label{sec:baseline}
We apply a multi-class classification approach to label the articles of the sampled dataset 
 into the six WikiProject quality classes. 
In order to have a baseline, we first apply the state of the art model proposed in~\cite{Wang2013} to the dataset. 
 

The ``actionable model" in~\cite{Wang2013} focuses on five
linguistic and structural features and it weighs them as follows:
\begin{enumerate}
\item Completeness = 0.4*{\it NumBrokenWikilinks} + 0.4*{\it NumWikilinks}
\item Informativeness = 0.6*{\it InfoNoise} + 0.3*{\it NumImages}
\item {\it NumHeadings}
\item {\it ArticleLength}
\item {\it NumReferences}/{\it ArticleLength}
\end{enumerate}
where 

\begin{itemize}
\item {\it NumWikilinks} is
the number of links pointing to other Wikipedia pages (whereas {\it NumBrokenWikilinks} counts the links that are broken); 
\item {\it InfoNoise} is the proportion of text content remaining in the article after removing MediaWiki markups and cleaning the text with basic NLP operations, such as stopwords removal;
\item {\it ArticleLength} is the base 10 log of the article length in bytes; 
\item {\it NumHeadings}, {\it NumReferences} and  {\it NumImages} are, quite intuitively, the number of headers, references and  images that an article contains. 
\end{itemize}

In order to evaluate such model over our dataset, we have extracted from our the dataset the above mentioned features. We have measured {\it ArticleLength},  {\it NumWikilinks} and {\it NumBrokenWikilinks} as suggested in~\cite{DelaCalzada}.

As shown in Figure~\ref{fig:schema}, the actionable model features have been extracted applying simple scripts mainly based on regular expressions
(the ``regexp extractor" block), which process the whole HTML code of an article (free text plus media wiki tags). The regexp extractor relies on
Python {\it BeautifulSoup}\footnote{\url{http://www.crummy.com/software/BeautifulSoup/}} for 
extracting HTML structures and  excerpts of  the textual content within the MediaWiki tags and on \textit{nltk} 
libraries\footnote{\url{http://www.nltk.org/}} for basic NLP analysis. In details, nltk has been used for computing the
InfoNoise feature, whose computation includes the stopwords removal, following the Porter Stopwords Corpus available through nltk~\cite{BirdKleinLoper09}. 

The classification results according to the baseline model are reported in Section~\ref{sec:exp}.
 
\section{The medical domain model}\label{sec:medical}
\begin{figure}
\centering
\includegraphics[scale=0.4]{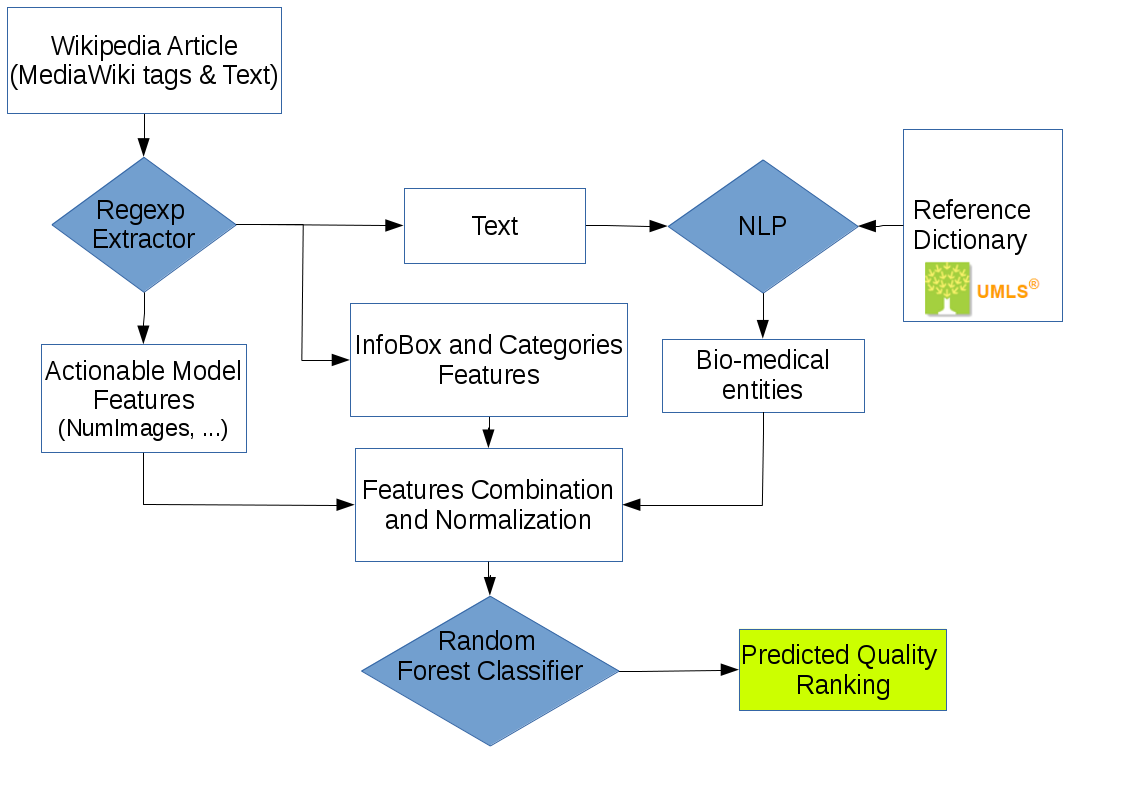}
\caption{Quality Assessment}
\label{fig:schema}
\end{figure}

Here, we improve the baseline model with novel and specifically crafted
features that rely on the medical domain and that capture details on the specific content of an article.
As shown in Figure \ref{fig:schema}, medical model features, the bio-medical entities, have been extracted 
from the free text only, exploiting advanced NLP techniques and using domain dictionaries.

In details, we newly define and extract from the dataset the following novel features: 
\begin{enumerate}
\item {\it InfoBoxNormSize}: this feature represents the normalised size of an infobox that contains  standard medical coding.
\item {\it Category}: the category a page belongs to.
\item {\it DomainInformativeness}: the number of bio-medical entities, which are 
 the domain dependent terms in the article (such as the ones denoting symptoms, diseases, treatments, etc.).
\end{enumerate}

The idea of considering infoboxes is not novel: for example, in~\cite{Wang2013} the authors noticed that the presence of an infobox  is a characteristic featured by good articles. 
However, in the specific case of the Medicine Portal, the presence of an infobox does not seem strictly related to the quality class the article belongs to (according to the manual labelling). 
Indeed, it is recurrent that articles, spanning all classes, have an infobox, containing 
a schematic synthesis  of the article. In particular, pages with descriptions of diseases usually have an infobox with  the medical standard code of the disease (i.e., IDC-9 and IDC-10), as in Figure~\ref{fig:infobox}. 

As done for the baseline, also the first two features of the medical model have been extracted with ad hoc Python scripts, extracting HTML structures and excerpts
of  the textual content within the MediaWiki tags. 

For their extraction of the bio-medical entities, we consider
the textual part of the article only, obtained after removing the MediaWiki tags, and we apply a NLP analysis, which is presented in Section~\ref{sec:nlp}.


\subsection{Infobox-based feature}
We have calculated the Infobox size as the base 10 log of the bytes of data contained within the mediawiki tags that wrap an infobox, and we have normalized it with respect to the ArticleLength, introduced in Section~\ref{sec:baseline}.

\subsection{Category-based feature}
We have leveraged the categories assigned to articles in Wikipedia, in particular relating to the medicine topics available at \url{https://en.wikipedia.org/wiki/Portal:Medicine}.
\begin{table}[htb]
\centering
\begin{tabular}{lr}
\toprule
category&list of keywords\\
\midrule
A&anatom*, embryolog*, organ, tissue\\
B&born, death, birth\\
D&disorder, disease, pathology\\
F&first aid\\
\bottomrule
\end{tabular}
\caption{Categories\label{tcat}}
\end{table}

We have defined 4 upper level categories of our interest:
\begin{itemize}
\item A {\it anatomy}:  an article is about anatomy;
\item B {\it biography}:  an article is a biography of someone or tell the history of something;
\item D {\it disorder}: it is about a disorder;
\item F  {\it first aid}: it reports information for first aid or emergency contacts;
\item O {\it other}: none of the above.
\end{itemize}

We have matched the article's  text within the MediaWiki categories tag with an approximate list of keywords that are related to our category of interest, as reported in Table \ref{tcat}.


\section{Bio-medical entities}\label{sec:nlp}
In the literature, there are several methods available for extracting bio-medical entities from a text (i.e., from medical notes and/or articles). 
We refer to~\cite{SemevalProc} for an overview of valuable existing techniques. 
In this work,  we have adopted a dictionary-based approach, which exploits lexical features and domain knowledge extracted from the Unified Medical Languages System (UMLS) Metathesaurus~\cite{Bodenreider2003}. The approach has been proposed for the Italian language in a past work~\cite{attardi2014adapting}. 
Since the approach combines the usage of linguistic analysis and domain resources, we were able to conveniently adapt it for the English language, being  both the linguistic pipeline and UMLS available for multiple languages (including English and Italian). 


Dictionary-based approaches have been proved valid for the task of entities' extraction, see, for example, another well known, similar approach to the one adopted here, i.e., Metamap\footnote{\url{http://metamap.nlm.nih.gov/}}. It is worth noting like, even though dictionary-based approaches could be less precise than  Named Entity Recognition~\cite{SemevalProc}, in our context even an approximate solution is enough, since we are not annotating medical records. Instead, we are quantifying the mole of inherent information within a text. 

\subsection{Reference dictionary}\label{sec:dictionary}
Several ontologies or taxonomies related to the medical domain are available in English.
To build a medical dictionary, we have extracted definitions of medical entities from the Unified Medical Languages System (UMLS) Metathesaurus~\cite{Bodenreider2003}.
UMLS integrates bio-medical resources, such as SNOMED-CT\footnote{\url{http://www.ihtsdo.org/snomed-ct/}} that provides the core terminology for electronic health records. In addition, UMLS also provides a semantic network where each entity in the Metathesaurus has an assigned Concept Unique Identifier (CUI) and it is semantically typed. 

From UMLS, we have extracted the entries  belonging to the following SNOMED-CT semantic groups: {\it Treatment}, {\it Sign or Symptom}, {\it Disease or Syndrome}, {\it Body Parts, Organs, or Organ Components}, {\it Pathologic Function}, and {\it Mental or Behavioral Dysfunction}, for a total of more than one million entries, as shown in Table~\ref{t2} (where the two last semantic groups have been grouped together, under {\it Disorder}).
Furthermore, we have extracted common Drugs and Active Ingredients definitions from RxNorm\footnote{\url{https://www.nlm.nih.gov/research/umls/rxnorm/}}, accessed by RxTerm\footnote{\url{https://wwwcf.nlm.nih.gov/umlslicense/rxtermApp/rxTerm.cfm}}.
\begin{table}[htb]
\centering
\begin{tabular}{lr}
\toprule
semantic groups&definitions\\
\midrule
Treatment&671,349\\
Sign or Symptom&43,779\\
Body Parts, Organs, or Organ Components&234,075\\
Disorder&402,298\\
Drugs&5,109\\
Active Ingredients&2,774\\
\bottomrule
\end{tabular}
\caption{Dictionary Composition\label{t2}}
\end{table}

Starting from the entries in Table~\ref{t2}, we have also computed approximate definitions, exploiting syntactic information of the same entries. 
In details, 
we have
pre-processed the entries by mean of the \textit{Tanl
  pipeline}~\cite{tanl-pipeline}, a suite of modules for text
analytics and NLP, based on machine learning.
Preprocessing has consisted in first dividing the entries into single word forms. Then, for each form, we have identified the lemma (when
available) and the part of speech (POS). Thus, we have created an approximate definition that consists in using only the lemma and cleaning the text, excluding punctuation, prepositions and articles. Also, approximate definitions have been normalized by lowercasing each word.
As an example, the Disorder entry ``aneurysm of the vein of galen" has been stored in the dictionary, along with its approximate definition ``aneurysm vein galen". 

\subsection{Extraction of bio-medical entities}
We have extracted the bio-medical entities present in the Wikipedia medical articles through a n-gram-based technique.

\begin{figure}[ht]
\centering
\includegraphics[scale=0.8]{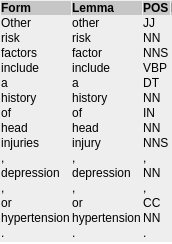}
\caption{Annotation of a sentence with the Tanl English pipeline\label{fig:postag}}
\end{figure}

A pre-processing phase occurs  in a similar way as for the dictionary composition. 
Given a Wikipedia article written in English, we have pre-processed the textual part through the \textit{Tanl
  pipeline}.
Similar to what described in Section~\ref{sec:dictionary} for the reference dictionary, we have first  divided the text in sentences and  the sentences into single word forms. For each form, we have considered the lemma (when
available) and the part of speech (POS).  For instance, starting from
an example sentence extracted from the Wikipedia page on the Alzheimer's disease:  ``{\it Other risk factors include a history of head injuries, depression, or hypertension.}",  we have obtained the annotation shown in
Figure~\ref{fig:postag}.
As in the case of the dictionary, each word in the text has been lowercasing.  

After pre-processing the text of each article, we have attempted to match each n-gram (with n between 1 and 10) in the corpus with the entries in the extended dictionary.
We both attempt an exact match and  an approximate match, the latter removing prepositions, punctuations and articles from the n-grams. Approximate matching leads to several advantages. 
Indeed,  exploiting the text pre-processing allows to identify dictionary definitions present  in the text, even when the number differs. As an example, the dictionary definition ``injury" will match with ``injuries", mentioned in the text, because in the approximation one can consider the lemmas. Further, considering the POS allows to identify mentions when interleaved by prepositions, articles, and conjunctions  that change the form but do not alter the meaning. As an example, the approximate definition ``aneurysm vein galen'' will match also with the following n-gram: ``the aneurysm and vein of galen", if present in the text.

\section{Experiments and results}\label{sec:exp}
In this section, we describe the experiments and report the results for the classification of
Wikipedia medical articles into the six classes of
the Wikipedia Medicine Portal. 
We compare the results obtained adopting four
different classifiers: the actionable model in~\cite{Wang2013} and
three classifiers that leverage the ad-hoc features from the medical
domain discussed in the previous sections.  All the experiments were
realized within the Weka framework~\cite{weka} and validated through
10 fold cross-validation.

For each experiment, we relied on the dataset presented in
Section~\ref{sec:dataset}, and specifically, on that obtained after sampling the majority classes and oversampling the minority ones (right-hand column in Table~\ref{table:t1}). The dataset serves both as training and test set for the classifiers. 

Moreover, to take into account the imbalanced data, we have applied
several classification algorithms and, for the sake of conciseness,
hereafter we report only the best results we have achieved. In
particular, we have experimented with bagging, adaptive boosting and
random forest and we report the results for the latter only.

\subsection{Classifiers' features}
\begin{table}[t]
\centering
\hspace{-0.2cm}
\setlength{\tabcolsep}{12pt}
\begin{tabular}{lllc}
\toprule
~		& ~		& \multicolumn{1}{c}{\textbf{Full}}& ~\\
\textbf{Baseline} & \textbf{Medical Domain} & \textbf{Medical Domain}& \textbf{Info Gain}\\
\midrule
ArticleLength & ArticleLength    & ArticleLength&  0.939\\
NumHeadings 	& NumHeadings      & NumHeadings&  0.732\\
Completeness & Completeness     & Completeness 	&  0.724\\
NumRef/Length 	& NumRef/Length    & NumRef/Length 	& 0.621\\
Informativeness	& Informativeness & Informativeness 	& 0.377\\
~ 		& DomainInformativ.	& DomainInformativ. &0.751\\
~ 		& ~  		  & InfoBoxNormSize 	& 0.187\\
~ 		& ~  		  & Category 		& 0.017\\
\bottomrule
\end{tabular}
\caption{Classifiers: Features and Information Gain\label{tab:features}}
\end{table}

 In Table~\ref{tab:features},
we report a summary of the  features for each of the considered
models: the baseline model in~\cite{Wang2013} and two new models that employ the medical domain features.   In the \textit{Medical Domain} model, we add to the baseline features the  Domain Informativeness, as described in
Section~\ref{sec:medical} and \ref{sec:nlp}. In addition, the \textit{Full Medical Domain} model also considers
 the  features InfoBoxNormSize and Category. 
 
For each of the features, the table also reports the Information Gain,
evaluated on the whole dataset (24,362 articles). Information Gain is a well-known 
metric to evaluate the dependency of one class from a single feature, see, e.g.,~\cite{Cover2006}. 

We can observe how  the
Domain Informativeness feature has a considerably higher infogain value when compared
with Informativeness. We anticipate here that this will lead to a more accurate
classification results for the highest classes, as reported in the next
section. Leading to a greater accuracy is also true for the other two new features that, despite
showing lower values of infogain, are able to further improve
the classification results, mainly for the articles belonging to 
the lowest quality classes (Stub and Start).


\subsection{Classification results}




Table~\ref{tab:comparison} shows the results of our multi-class
classification. For each of the classes, we have computed the
\textit{ROC Area} and \textit{F-Measure}
metrics~\cite{powers2011evaluation}. The latter, in particular, is
usually considered a significant metric in terms of classification,
since it combines in one single value all the four indicators that are
generally implied for evaluating the classifier performance (i.e.,
number of True Positives, False Positives, True Negatives and False
Negatives). In our scenario, the meaning of the indicators, for each
class, are as follows:
\begin{itemize}
\item \textit{True Positives} are the articles classified as belonging to a certain class, that indeed belong to that class (according to the quality ratings given by the WikiProject Medicine);
\item \textit{True Negatives} are the articles classified as not belonging to a certain class, that indeed do not belong to that class;
\item \textit{False Positives} are the  articles classified as belonging to a certain class, that  do not belong to that class;
\item \textit{False Negatives } are the  articles classified as not belonging to a certain class, that instead belong to that class. 
\end{itemize}

\begin{table}[t]
\centering
\setlength{\tabcolsep}{12pt}
\begin{tabular}{lccc}
\toprule
~ & ~		& \textbf{Medical} & \multicolumn{1}{c}{\textbf{Full Medical}}\\
\textbf{Metric} & \textbf{Baseline} & \textbf{Domain} & \textbf{Domain}\\
\midrule
ROC Area Stub	&0.981	&0.982	&\textbf{0.983}\\
ROC Area Start	&0.852	&0.853	&\textbf{0.858}\\
ROC Area C	&0.749	&0.747	&\textbf{0.76 }\\
ROC Area B	&0.825	&0.832	&\textbf{0.836}\\
ROC Area GA	&0.825	&0.908	&\textbf{0.916}\\
ROC Area FA	&0.977	&0.976	&\textbf{0.978}\\
\midrule
F-Measure Stub	&0.886	&\textbf{0.891}	&0.89 \\
F-Measure Start	&0.587	&0.582	&\textbf{0.598}\\
F-Measure C	&0.376	&0.367	&\textbf{0.397}\\
F-Measure B	&0.527	&0.541	&\textbf{0.542}\\
F-Measure GA	&0.245	&0.338	&\textbf{0.398}\\
F-Measure FA	&0.634	&0.631	&\textbf{0.641}\\
\bottomrule
\end{tabular}
\caption{Classification Results\label{tab:comparison}. In bold, the best results.\label{tab:comparison}}
\end{table}

At a first glance, we observe that, across all the models, the
articles with the lowest classification values, for both ROC and F-Measure, are those labeled  C and GA. Adding the Domain Informativeness feature produces a classification, which is  slightly worse for  C and FA
articles, but better for the other four classes. This is particularly evident for the
F-Measure of the articles of the GA class. 
A noticeable major
improvement is obtained with the introduction of the
features InfoBoxNormSize and Category in the \textit{Medical Domain} model.
%
The ROC Area increases for the articles of all the classes within the
\textit{Full Medical Domain}, while the F-Measure is always better than the \textit{Baseline} and almost always better than the \textit{Medical Domain}.


The size of an article, expressed either as the word count, analyzed in~\cite{Blumenstock2008}, or as the article length, as done here, appears a very strong feature, able to
discriminate the articles belonging to the highest and lowest quality classes. This is testified also by the results achieved exploiting the baseline model 
of~\cite{Wang2013}, which poorly succeeds in discriminating the
articles of the intermediate quality classes, while achieving  good results for Stub
and FA.  
Here, the newly introduced features have a predominant effect on
the articles of the highest classes. This could be justified by the
fact that those articles contain, on average, more text and, then, NLP-based
features can exploit more words belonging to a specific
domain. 

Then, we observe that the ROC Area and the F-Measure are
not tightly coupled (namely: high values for the first metric can
correspond to low values for the second one, see for example C and
GA): this is due to the nature of the ROC Area, that is affected by
the different sizes of the considered classes. As an example, we can
observe that the baseline model has the same ROC Area value
for the articles of both class B and class GA, while the F-Measure of
articles of class B is 0.282 higher than that of class GA. 

Finally,
the results confirm that the adoption of domain-based features and, in
general, of features that leverage NLP, help to distinguish between
articles in the lowest classes and articles in the highest classes, as highlighted in bold
in Table~\ref{tab:comparison}. We notice also that exploiting the full medical domain leads us to the achievement of the best results. 

Even if preliminary, we believe that the results are promising and call 
both for features' further refinement and novel features, 
able to discriminate among the intermediate classes too.



\section{Related work}\label{sec:RW}

Automatic quality evaluation of Wikipedia articles has been addressed in
previous works with both unsupervised and supervised learning approaches.
The common idea of most of the existing work is to identify a feature set, having
as a starting point the Wikipedia project guidelines, to be exploited with the objective in mind to distinguish Featured Articles.
In~\cite{stvilia2005}, Stvilia \textit{et al.} identify a relevant
set of features, including lingual, structural, historical and reputational aspects of each article. They show the
effectiveness of their metrics by applying both clustering and
classification. 
As a result, more than 90\% of FA are
correctly identified.

Blumenstock~\cite{Blumenstock2008} inspects the relevance of the \textit{word-count} feature at each quality stage, showing that it can play a very important role in the quality assessment of Wikipedia articles. Only using this feature, the author achieves a F-measure of 0.902 in the task of classifying featured articles  and 0.983 in the task of classifying non featured articles. The best results of the investigation are achieved with a classifier based on a neural network implemented with a multi-layer perceptron.

In~\cite{WuZhuZhao}, the authors try to analyze the factors affecting the quality of Wikipedia articles, with respect to their quality class. The authors
evaluate a set of 28 features, over a random sample of 500 Wikipedia
articles, by weighing each metric in different stages  using neural networks. Findings are that linguistic features weigh more in the lowest quality classes, and structural features, along with historical ones, become more important as the article quality improves. 
Their results indicate that the information quality is mainly affected by completeness, and to be ``well-written" is a basic requirement in the initial stage. Instead, reputation of authors or editors is not so important in Wikipedia because of its horizontal structure.
In~\cite{Witold15}, the authors consider  the quality of the data in the  infoboxes of Wikipedia, finding  a correlation between the quality of information in the infobox and the article itself.

In~\cite{Wang2013}, the authors deal with the problem of discriminating
between two large classes, namely \emph{NeedWork, GoodEnough} (including in GoodEnough both GA and FA), in order to identify
which articles need further revisions for being featured.  They also
introduce new composite features, those that we have referred to as an ``actionable model'' in Section~\ref{sec:baseline}. 
They obtain  good classification results, with a F-measure of 0.876 in their best configuration.  They
also try classification for all the seven quality classes, as done in this work, using a
random forest classifier with 100 trees, with a reduced set of
features. The poor results (an average F-measure of 0.425) highlights
the hardness of this fine-grained classification.
In this paper, we address this last task in a novel way, by introducing domain features, specially dealing with the medical domain. The results of the investigation are promising.

Recent studies specifically address the quality of medical information (in  Wikipedia as well as in other resources): in~\cite{Azer5} and ~\cite{Azer6}, the authors debate if Wikipedia is a reliable learning resource for medical students, evaluating articles on respiratory topics and cardiovascular diseases. The evaluation is carried out by exploiting DISCERN\footnote{\url{http://www.discern.org.uk/}}, a tool evaluating readability of articles. 
In~\cite{odbase2014} the authors provide novel solutions for measure the quality of medical information in Wikipedia, by adopting an unsupervised approach based on the Analytic Hierachy Process, a multi-criteria decision making technique~\cite{ahp90}. 
The work in~\cite{Cabtiza13}  aims to provide the web surfers a numerical indication of Quality of Medical Web Sites. 
In particular in~\cite{Cabtiza13} the author proposes an index to make IQ judgment of the content and of its reliability, to give the so called ``surface markers'' and ``trust indicator''. A similar measurement is considered in~\cite{SI:ASI21115}. where  the authors present an empirical analysis that suggests the need to define genre-specific templates for quality evaluation and  to develop models for an automatic genre-based classification of health information Web pages. In addition, the study shows that consumers may lack the motivation or literacy skills to evaluate the information quality of health Web pages. Clearly, this further highlights the cruciality to develop accessible automatic information quality evaluation tools and ontologies. Our work moves towards the goal, by specifically considering domain-relevant features and featuring an automatic classification task spanning over more than two classes.  

\section{Conclusions}\label{sec:concl}
In this work, we aimed to provide a fine grained classification
mechanism for all the  quality classes of the articles of the
Wikipedia Medical Portal. The idea was to propose an automatic
instrument for helping the reviewers to understand which articles are
the less work-demanding papers to pass to next quality stage. We focused
on an actionable model, namely whose features are
related to the content of the articles, so that they can also directly
suggest strategies for improving a given article. An important and
novel aspect of our classifier, with respect to previous works, is the
leveraging of features extracted from the specific, medical domain,
with the help of Natural Language Processing techniques. As the
results of our experiments confirm, considering specific domain-based
features, like Domain Informativeness and Category, can eventually
help and improve the automatic classification results. Since the results are
encouraging, as future work we will evaluate other
features based on the specific medical domain. Moreover, we are
planning to extend our idea, to include and compare also other
non medical articles (thus, extending the work to include other domains), in order to further validate our approach.

\noindent\textit{\bf Acknowledgments.}
The research leading to these results has been partially funded by the Registro.it project My Information Bubble MIB.
%
\bibliographystyle{abbrv}
\bibliography{biblio}

\end{document}